\documentclass[onecolumn]{emulateapj}

\newcommand{\cm}{{\rm cm}}
\newcommand{\s}{{\rm s}}
\newcommand{\kms}{{\rm km}\,{\rm s}^{-1}}

\newcommand{\kpc}{{\rm kpc}}
\newcommand{\Mpc}{{\rm Mpc}}
\newcommand{\erg}{{\rm erg}}
\newcommand{\yr}{{\rm yr}}
\newcommand{\Hz}{{\rm Hz}}
\newcommand{\Ryd}{{\rm Ryd}}

\newcommand{\HI}{\ion{H}{1}} 
\newcommand{\HeII}{\ion{He}{2}} 
\newcommand{\CII}{\ion{C}{2}}
\newcommand{\CIV}{\ion{C}{4}}
\newcommand{\OVI}{\ion{O}{6}}
\newcommand{\SiIV}{\ion{Si}{4}}
\newcommand{\MgII}{\ion{Mg}{2}}

\newcommand{\lya}{Ly$\alpha$}

\shorttitle{Local radiation and QSO absorbers}
\shortauthors{Schaye}

\begin{document}

\slugcomment{Accepted for publication in the Astrophysical Journal}

\title{On the importance of local sources of radiation for\\
  quasar absorption line systems}  

%\author{Joop Schaye\altaffilmark{1}} 
%\altaffiltext{2}{School of Natural Sciences, Institute for Advanced
%Study, Einstein Drive, Princeton NJ 08540; schaye@ias.edu}
\author{Joop Schaye} 
\affil{Leiden Observatory, P.O. Box 9513, 2300 RA Leiden, The
  Netherlands, schaye@strw.leidenuniv.nl} 
\affil{School of Natural Sciences, Institute for Advanced
Study, Einstein Drive, Princeton NJ 08540}

\begin{abstract}
A generic assumption of ionization models of quasar absorption systems
is that radiation from local sources is negligible compared with the
cosmological background. We test this assumption and find that it is
unlikely to hold for absorbers as rare as \HI\ Lyman limit systems.
Assuming that the absorption systems are gas clouds centered on
sources of radiation, we derive analytic estimates for the
cross-section weighted moments of the flux seen by the absorbers, of
the impact parameter, and of the luminosity of the central source. In
addition, we compute the corresponding medians numerically.  For the
one class of absorbers for which the flux has been measured: damped
\lya\ systems at $z\approx 3$, our prediction is in excellent
agreement with the observations if we assume that the absorption
arises in clouds centered on Lyman-break galaxies. Finally, we show
that if
Lyman-Break galaxies dominate the UV background at redshift 3, then
consistency between observations of the UV background, the UV 
luminosity density from galaxies, and the number density of Lyman
limit systems requires escape fractions of order 10 percent.
\end{abstract}


\section{Introduction}
The intergalactic medium (IGM) is thought to be photo-ionized by the UV
radiation emitted by galaxies and quasars. Models of quasar absorption
systems generally assume that the gas is exposed only to the mean
background radiation that pervades the IGM. However, given that this
background is produced by discrete sources, it cannot be
strictly uniform. 

The effect of fluctuations in the UV background on the statistics of
the forest of weak HI \lya\ absorption lines, seen in the spectra of
distant quasars, has been the subject of a large number of papers
(e.g., Zuo 1992; Fardal \& Shull 1993; Croft et al.\ 1999, 2002;
Gnedin \& Hamilton 2002; Linder et al.\ 2003; Meiksin \& White 2003,
2004; Croft 2004; McDonald et al.\ 2005). These theoretical studies
found that because most of the 
\lya\ lines arise in the low-density IGM, far away from sources of
ionizing radiation, global statistics of the forest are
insensitive to fluctuations in the UV background. However,
fluctuations in the \HI\ ionization rate may become detectable in the
\lya\ forest at at $z>5$ (e.g., Meiksin \& White 2003, 2004) and
fluctuations in the \HeII\ 
ionization rate have already been detected at $z\ga 2.5$ from
comparisons of the \HI\ and \HeII\ \lya\ forests (e.g., Kriss et al.\
2001; Smette et al.\ 2002; Shull et al.\ 2004).  

In contrast to the \lya\ forest, many metal and high column
density \lya\ absorbers are
thought to arise in the extended halos of galaxies (Bahcall \& Spitzer
1969), or at least in regions close to galaxies. It is therefore not
obvious that neglecting the contributions of local sources of 
ionizing radiation,
which is currently common practice, is justified for
such absorption systems. Indeed, it has been argued on the basis of
metal-line column density ratios that there is evidence that some
absorbers are exposed to a radiation field that is much softer than the
general background, as would be expected if local stellar radiation
were important (e.g., Giroux \& Shull 1997; Boksenberg, Sargent, \&
Rauch 2003). 

Here we show that if we assume that a certain class of absorbers
resides in the extended halos\footnote{By ``halo'' we mean a roughly
  spherical region centered on a source. It does not need to be
  virialized, or even be gravitationally bound.} of a population of
sources of UV 
radiation (such as galaxies), then \emph{one can estimate the typical
flux from local sources to which the absorbers are exposed using only
the luminosity density of the sources and the rate of incidence (i.e.,
number per unit redshift) of the absorbers.} We do not need to know the
relation between the cross-section for absorption and the luminosity
of the source.

We derive analytic formulas to estimate the moments of the local flux,
as well as of the 
impact parameter and the luminosity of the central source, and apply them
to galaxies and quasars. We find that UV
radiation from local galaxies may well be important for
absorbers rarer than Lyman limit (LL) systems (in agreement with the
recent work of Miralda-Escud\'e 2005) and is likely far in excess
of the background for absorbers as rare as damped Ly$\alpha$ (DLA)
systems. We therefore
conclude that the results from studies that employed
ionization models of $N_{\rm HI} > 10^{17}\,\cm^{-2}$ absorbers may need to be
revised. 

This paper is organized as follows. In \S\ref{sec:assumptions} we
state and discuss the approximations and assumptions which we use to
estimate the local flux. Section \ref{sec:meanflux}
contains the derivation of the mean flux, which we show to be close to
the median for a wide range of models in \S\ref{sec:medians}. Section
\ref{sec:comp} presents two complementary methods to compare the local
flux to the background flux and contains some general
conclusions. Using
more restrictive assumptions, we derive expressions for the moments of
the cross-section weighted flux, impact parameter, and luminosity in
\S\ref{sec:moments}. In 
\S\S\ref{sec:gals} and \ref{sec:quasars} we compute the contributions
of galaxies and quasars to the local \HI\ ionization rate and we
compare these with the background for various types of
absorbers in \S\ref{sec:comparison}, where we also estimate the global
escape fraction for H ionizing photons. Finally, we summarize our main
conclusions in \S\ref{sec:conclusions}.

\section{Estimating the flux}
\label{sec:method}
It is clearly beyond our means to compute the distribution of fluxes
seen by a class of absorbers from first principles. Such a calculation
would involve modeling radiative transfer in a characteristic
volume of the universe, which would require specification of the 
full phase space distribution of 
elements and sources as well as of the
spectra emitted by the sources. We would then still need to decide
what we mean by \emph{the} 
flux seen by an absorber as the absorbing gas clouds have a finite
size. It is therefore necessary to make a number of simplifying
assumptions. We will present and discuss these assumptions in
\S\ref{sec:assumptions} and estimate the mean flux in
\S\ref{sec:meanflux}. In \S\ref{sec:medians} we will show that the
median flux, which we can only compute by making additional
assumptions, is generally close to the mean. 

\subsection{Model assumptions}
\label{sec:assumptions}
Consider a certain class of absorption systems with an observed rate
of incidence $d{\cal N}/dz$. In practice, we will define a class of
absorbers by specifying an ion and a minimum column density $N_{\rm
min}$.  To enable us to compute the flux from local sources (e.g.,
galaxies) to which
the absorbers are exposed, we will make the following simplifying
assumptions and approximations regarding absorbers with $N>N_{\rm min}$: 
\begin{enumerate} 
\item All absorbers reside in spherical halos which are each centered
around a single source.
\item The flux seen by the absorber is dominated by its central
source.
\item The probability that a sightline with impact parameter $b$
relative to a source with luminosity $L$ intersects an absorber
residing in the
halo around the source is $P_{\rm abs} = f_{\rm cov}$ for $b\le R(L,N_{\rm
  min})$ and zero otherwise 
[i.e., the absorbing halo has a finite radius $R(L,N_{\rm min})$].
\item All of the gas contributing to the absorption resides at a
distance $R(L,N_{\rm min})$ from the central source.
\item The central source is point-like.
\item The product $f_{\rm cov}f_{{\rm esc},N}$, where $f_{{\rm esc},N}$ is the
fraction of the emitted radiation that is able to propagate to the
absorber, is independent of the luminosity of the source.
\end{enumerate}
We will now discuss each of these assumptions in turn.

The assumption that the absorbers are (or reside in) spherical halos
centered on single sources will clearly break down 
for absorbers with gas densities around the cosmic mean, as these are
typically distributed along sheets and filaments. However, it is
probably a reasonable approximation for
gas with densities $\ga 10$ times the cosmic mean and/or
for heavy elements whose distributions are likely concentrated
towards galaxies. 

The second assumption (radiation from sources in nearby halos is
negligible) is conservative. We will show in \S\ref{sec:gals} that it is
likely to be a good approximation if the sources are galaxies and if
the flux from the central source exceeds the
background. However, if
the sources are strongly clustered, then radiation from nearby 
halos could provide a non-negligible contribution to the ionizing flux
seen by the absorbers. 

Assumption three ($P_{\rm abs} = 0$ for $b>R$) is probably reasonable
provided that we define a class by its cumulative rate of incidence
$d{\cal N}/dz(N>N_{\rm min})$. 
Since $d{\cal N}/dz(N>N_{\rm min})$ must decrease with increasing
$N_{\rm min}$, the assumption implies
that larger columns typically arise in sightlines
with smaller impact parameters and hence that $f(N,z) \equiv d^2{\cal
N}/dzdN$ is a decreasing function of $N$. Although it is possible to
fabricate density distributions that would violate this assumption, it
cannot be violated systematically because that would mean that the
cosmic density of the ion in question, which is proportional to the
integral $\int N f(N,z)dN$, would diverge at large N. Moreover, the
assumption that column density typically increases with decreasing
impact parameter is supported by a large number of observational
studies of the relation between absorption systems and galaxies (e.g.,
Lanzetta \& Bowen 1990; Bergeron \& Boisse 1991; Lanzetta et al.\
1995; Chen et al.\ 1998, 2001a, 2001b; Bowen, Pettini, \& Blades 2002;
Penton, Stocke, \& Shull 2002; Adelberger et al.\ 2003).

One might naively think that the assumption that column density
increases with decreasing impact parameter would break down in the
region within which the ionizing flux is
dominated by the central source (i.e., the proximity zone). This would
be a problem, as 
this is one of the regimes of interest here. However, the arguments
given above indicate that this cannot generally be the case. Before
discussing our remaining assumptions, we will show that this is not a
paradox. 

Although the neutral hydrogen column density in a sightline passing
close to a source can be small if the proximity zone is very large (as
can be the case for bright quasars), the column density is unlikely to
increase with the impact parameter for any given source. To see this,
consider that in photo-ionization equilibrium the ionization balance
is determined by the ratio of the ionizing flux to the gas density,
which is independent of the radius if the flux is dominated by the
central source and if the gas density profile is isothermal, which
should be close to the situation of interest here. In general,
for an optically thin gas in photo-ionization equilibrium that is
irradiated by a central source we have $n_{\rm HI}
\propto n_{\rm H}^2 \Gamma_{\rm HI}^{-1} \propto
r^{-a}$ if $n_{\rm H} \propto r^{-(2+a)/2}$, where $\Gamma_{\rm
  HI}\propto r^{-2}$
is the photo-ionization rate for neutral hydrogen. Now consider a
spherical cloud with density profile $n\propto r^{-a}$, where $n$ is the number
density of the species of interest. The column
density in a sightline with impact parameter $b$ is
\begin{equation}
N(b) \propto \int {l \over (b^2 + l^2)^{a/2}} d\ln l.
\label{eq:N(b)}
\end{equation}
For $a>1$ the dominant contribution to the column density comes
from $l\la b$ and $N\sim n(b)b$. For example, for a singular
isothermal sphere ($a=2$) we have $N=\pi n(b)b \propto b^{-1}$, which
is the scaling derived by Schaye
(2001a, equations 3 and 4) based on more general arguments.
Hence, for
reasonable density profiles the column density will increase with
decreasing impact parameter.

There are two reasons why we would in general expect this trend to be
even stronger for metals than for \HI. First, since metals are
produced in stars, it is natural to assume that their abundances
relative to hydrogen are decreasing functions of $R$. Second, the
ionization balance of many 
of the metals of interest will asymptote to a
constant as more stellar radiation is added, because (ordinary) stars
essentially do not emit above 4~Rydberg (Ryd) and metals cannot be fully
photo-ionized 
without these energetic photons\footnote{Note that this implies that for
spectra with a sharp cutoff, such as those of galaxies, it is not
true that the ionization balance depends only on the ratio of hydrogen
ionizing photons to the gas density 
(i.e., the ionization parameter).}

Assumption four is also conservative: assuming that all the absorbing
gas resides in a thin shell with radius $R$ minimizes the flux from
the central source.  In fact, for a typical absorber it is probably a
reasonable approximation to assume that all absorbing gas is at the
maximum possible radius $R$. We already showed that for
reasonable density profiles almost all the absorption takes place in
the densest part of the gas cloud that is intersected by the
sightline, at radii slightly greater than $b$. The median impact 
parameter of random sightlines through a halo of radius $R$ is $\left
< b \right > = R/\sqrt{2}$, slightly smaller than $R$. Hence, the
column densities are typically dominated by gas at a distance $r\sim
R$ from the source. We therefore expect that the assumption that all
of the gas that contributes to the column density resides at a radius
$R$ will, for a typical absorber, result in an underestimate of the
flux seen by the absorber by only a factor of a few.

Note that because of assumption four, the flux seen by
absorbers in some finite column density interval
$N_{\rm min} \le N \le N_{\rm max}$ is identical to
that seen by absorbers with $N > N_{\rm min}$. 

Assumption five (point-like sources) will be a good approximation
for sightlines with $b \gg R_s$, where $R_s$ is the radius of the
source. For smaller impact
parameters it will, however, lead to a large overestimate of the
flux. Hence, our model should not be applied to absorbers with $R <
R_s$.

Finally, assumption six ($f_{\rm cov}f_{{\rm esc},N}$ is independent of
$L$) can be justified only by our ignorance. Given that we are still a
long way off 
from being able to compute this factor from first principles, it would
be hard to justify any particular dependence on $L$. Assuming it
is constant minimizes the number of free parameters.

The fraction of photons that is able to propagate to the absorber,
$f_{{\rm esc},N}$, should not be confused with the conventional escape
fraction $f_{\rm esc}$, which is the fraction of emitted photons
that is able to escape the entire halo surrounding the source. They
are different in general because the absorbers under consideration may
in fact determine the global escape fraction $f_{\rm esc}$. Thus, we
have $0 \le f_{\rm esc} \le f_{{\rm esc},N} \le 1$ in general with $f_{{\rm
    esc},N}$ asymptoting to $f_{\rm esc}$ for sufficiently small
$N_{\rm min}$ and to unity for sufficiently large $N_{\rm min}$ (i.e.,
for column densities comparable to those surrounding the sources). 

The covering factor $f_{\rm cov}$ ($0 < f_{\rm cov} \le 1$) is
unlikely to be much 
smaller than unity given that absorption line systems typically
contain multiple components and that sightlines which pass near to
galaxies nearly always show absorption by metals and hydrogen
(e.g., Bergeron \& Boisse 1991; Lanzetta et al.\ 1995; Adelberger et
al.\ 2003).

\subsection{The mean flux}
\label{sec:meanflux}
We are now in a position to estimate the flux from the central
source to which a typical absorber is exposed. We will compute the
mean flux $\bar{F}$ by averaging the flux seen by 
an absorber over all source luminosities, weighted by the total
cross-section for absorption. Under the assumptions
discussed above, a source of luminosity $L$ provides a (proper)
cross-section for absorption of $f_{\rm cov}\pi R^2$ and the flux seen
by the absorber is
\begin{equation}
F = {L f_{{\rm esc},N} \over 4\pi R^2},
\label{eq:F}
\end{equation}
The average number of absorption
systems per unit redshift and luminosity along a random sightline 
is
\begin{equation}
{d^2{\cal N} \over dzdL}(z,L,N_{\rm min}) = \Phi(z,L) f_{\rm cov} \pi
R^2(z,L,N_{\rm min}) {c \over H(z)} (1+z)^2, 
\label{eq:dN/dz}
\end{equation}
where $\Phi$ is the comoving luminosity function of the sources (i.e.,
the number of sources per unit comoving volume and per unit luminosity),
$c$ is the speed of light and $H$ is the Hubble parameter. 
The mean flux from the central source to which a class of
absorbers with rate of 
incidence $d{\cal N}/dz(z,N>N_{\rm min})$ is exposed is thus
\begin{eqnarray}
\bar{F} (z,N_{\rm min})
&=& \left [\int {d^2{\cal N} \over dz dL}dL
\right ]^{-1} \int {d^2{\cal N} \over dz dL} F dL \\
&=& \left [{d{\cal N} \over dz}(z,N_{\rm min})\right ]^{-1}{c (1+z)^2
  \over 4 H(z)} f_{{\rm esc},N} f_{\rm cov} \int L\Phi(z,L) dL.
\label{eq:barF}
\end{eqnarray}
The important thing to note is that the result is independent of 
$R$. Thus, \emph{we do not need to specify the function $R(L)$ in order to 
compute the mean flux $\bar{F}$}. Equation~(\ref{eq:barF}) is a key
result of this paper, it provides a means to 
estimate the flux seen by a class of absorbers from its observed
rate of incidence and the observed luminosity function of sources. 

The mean flux is proportional to the luminosity density $\int
L\Phi dL$. If the sources are galaxies, then this has the convenient
consequence that $\bar{F}$ is approximately proportional to the comoving star
formation density because most of the UV radiation is emitted by young
stars.

If the luminosity function is a Schechter (1976) function,
\begin{equation}
\Phi(L)dL = \phi_\ast\left ({L \over L_\ast}\right )^{\alpha}
e^{-L/L_\ast} {dL \over L_\ast},
\label{eq:lf}
\end{equation}
then equation (\ref{eq:barF}) becomes
\begin{equation} 
\bar{F} (z,N_{\rm min})
= \left [{d{\cal N} \over dz}\right ]^{-1}{c (1+z)^2
\over 4 H} f_{{\rm esc},N} f_{\rm cov} k \phi_\ast L_\ast \Gamma(2+\alpha), 
\label{eq:barF_sf}
\end{equation}
where $\Gamma(x)$ is the gamma function and $k = 1/(0.4\ln 10)$ if $\phi_\ast$
is given per magnitude and $k=1$ if $\phi_\ast$ is given per $\ln
L$ (unless specified otherwise, we will hereafter assume $k=1$ and
omit $k$ from the equations).
If, from some reason, the absorbers only reside in halos around
objects with $L_{\rm min} < L < L_{\rm max}$, then the integral should
of course be computed using these limits. 
For physically sensible luminosity
functions (i.e., those that yield a finite luminosity density), $\int
L \Phi dL \sim \Phi_{\ast} L_{\ast}$ and $\bar{F}$ will thus be
insensitive to the limits as long as $L_{\rm min} \ll L_\ast \ll
L_{\rm max}$.

\section{Comparing the local flux to the background}
\label{sec:comp}
To assess the importance of local sources, we need to compare the mean
flux from the central source to the background radiation field. If the
intensity of the background radiation is known, then we can directly
compare it to our estimate of the mean flux [equation (\ref{eq:barF})]
and infer the critical 
rate of incidence below which
local radiation dominates:
\begin{eqnarray}
\left ({d{\cal N} \over dz}\right )_{\rm crit}
&=& {d{\cal N} \over dz} {\bar{F} \over F_{\rm bg}} \\
\label{eq:dndzcrit0}
&=&
{1 \over F_{\rm bg}} {c (1+z)^2
  \over 4 H(z)} f_{{\rm esc},N} f_{\rm cov} \int L\Phi(z,L) dL,
\label{eq:dndzcrit1}
\end{eqnarray}
where $F_{\rm bg}$ is the background flux. For sufficiently large
$N_{\rm min}$ (i.e., small ${d{\cal N}/dz}$) $f_{{\rm esc},N}
\rightarrow 1$ and the critical rate of incidence can be determined
relatively accurately. But if $N_{\rm min}$ is much smaller than the
column in front of the sources, then $f_{{\rm esc},N} \rightarrow
f_{\rm esc}$ and the critical rate becomes proportional to the highly
uncertain global escape fraction.

On the other hand, if we assume that the population of sources under
consideration dominates the background, that the proper mean
free path of the photons $\lambda_{\rm mfp}$ is known, and that
$\lambda_{\rm mfp} \ll c/H(z)$ (such that cosmological redshift and
evolution may be neglected), 
then the background flux is given by
\begin{equation}
F_{\rm bg} = (1+z)^3 \lambda_{\rm mfp}(z) f_{\rm esc} \int L\Phi(z,L) dL.
\label{eq:fbg}
\end{equation}
Substituting this into equation (\ref{eq:dndzcrit1}), we see that the critical
rate of incidence becomes independent of 
both the source luminosity function and the intensity of the
background:
\begin{eqnarray}
\left ({d{\cal N} \over dz}\right )_{\rm crit} &=& 
{1 \over \lambda_{\rm mfp}(z)} {f_{{\rm esc},N} f_{\rm cov}
  \over f_{\rm esc}} {c \over 4 H(z) (1+z)}, \\
&=& {1 \over 4 z_{\rm mfp}(z)} {f_{{\rm esc},N} f_{\rm cov}
  \over f_{\rm esc}}, 
\label{eq:dndzcrit2}
\end{eqnarray}
where $z_{\rm mfp}$ is the redshift interval corresponding to the mean
free path.

For sufficiently large
$N_{\rm min}$, $f_{{\rm esc},N}
\rightarrow 1$ and the critical rate of incidence is inversely
proportional to the highly uncertain global escape fraction $f_{\rm
  esc}$. But for sufficiently small $N_{\rm min}$, 
$f_{{\rm esc},N} \rightarrow f_{\rm esc}$ and the critical rate of
incidence becomes independent of the escape fraction. 

Hence, for column densities similar to those in front of typical
sources, using equation (\ref{eq:dndzcrit1}) will probably give a more accurate
estimate, whereas using equation (\ref{eq:dndzcrit2}) will likely yield the
most accurate estimate
for lower column densities. In general, we can measure the global escape
fraction $f_{\rm esc}$ by comparing the two estimates, which 
comes down to comparing equation (\ref{eq:fbg}) with the measured
intensity of the background. Of course, such a measurement is subject
to the uncertainties in the luminosity density and in $\lambda_{\rm
  mfp}$. Moreover, if other sources contribute to the background, then
the inferred escape fraction should be interpreted as an upper limit.

Equation (\ref{eq:dndzcrit2}) corresponds with equation (4) of
Miralda-Escud\'e (2005) if we neglect evolution of the rate of
incidence on timescales $\la z_{\rm mfp}$, as we have done. However,
Miralda-Escud\'e (2005) implicitly assumed $f_{\rm cov}=1$ and $f_{{\rm
    esc},N} = f_{\rm esc}$. The last assumption may well break down for
column densities similar to those in front of the sources, such as
Damped Ly$\alpha$ (DLA) Systems. 

The mean free path depends on the frequency of the photons. For
energies between 1 and 4 Rydbergs (i.e., between the ionization
potentials of \HI\ and \HeII), the mean free path depends mainly on
the \HI\ column density distribution of absorbers with column
densities similar to those of LL systems  ($N_{\rm HI} > 1.6\times
10^{17}\,\cm^{-2}$). Unfortunately, this is
exactly where the distribution is most 
uncertain, because the \HI\ absorption lines are on the flat part of
the curve of growth which makes it difficult to measure accurate
column densities. Following Miralda-Escud\'e (2003), if we assume that
the column density distribution is a power-law $d{\cal
  N}/dN_{\rm HI} \propto N_{\rm HI}^{-\eta}$ with slope $\eta=1.5$, as
appears to be a reasonable fit at lower column densities (e.g., Kim et
al.\ 2002), then 
it can be shown that $\lambda_{\rm mfp} = \lambda_{\rm LL} /
\sqrt{\pi}$, where $\lambda_{\rm LL}$ is the mean spacing between
LL systems. This implies that $z_{\rm mfp}^{-1} = \left
(d{\cal N}/dz\right )_{\rm LL} \sqrt{\pi}$ and hence that the critical
rate of incidence is proportional to that of LL systems:
\begin{equation}
\left ({d{\cal N} \over dz}\right )_{\rm crit}
\sim {\sqrt{\pi} \over 4} \left ({d{\cal N} \over dz}\right )_{\rm LL}
{f_{{\rm esc},N} f_{\rm cov} \over f_{\rm esc}}, 
\label{eq:dndzcrit3}
\end{equation}
which is to be considered an order of magnitude estimate because it
makes strong assumptions about the shape of the column density
distribution. Note that this equation is less
general as the previous ones because it applies only to photons with
energy above (but near) 1~Ryd. 

If we assume that $f_{{\rm
    esc},N}f_{\rm cov} \sim f_{\rm esc}$, which is probably reasonable for
$N_{\rm HI} \la 10^{17}\,\cm^{-2}$, then we have $(d{\cal N}/
dz)_{\rm crit} \sim (d{\cal N} / dz)_{\rm LL}$, in agreement
    with Miralda-Escud\'e (2005). Thus, we
conclude that \emph{for hydrogen, local ionizing radiation is
generally unimportant for $N_{\rm HI} \ll 10^{17}\,\cm^{-2}$. For
Lyman limit systems local radiation is important, provided that the
    sources associated with these systems
dominate the UV background} (we expect this to be the case if the
background is dominated by stars, as opposed to quasars). 
The rate of incidence of DLAs is much smaller than that of Lyman limit
systems (about a factor 
10 smaller at $z=3$; Storrie-Lombardi et al.\ 1994; Storrie-Lombardi \&
    Wolfe 2000) which implies 
that \emph{local sources dominate the \HI\ ionization rate in
DLAs if the same population also dominates the UV background}. Note that
$\bar{F}$ may exceed $F_{\rm bg}$ by a very large 
factor if $f_{{\rm esc},N} \gg f_{\rm esc}$, as may well be the case
for DLAs.

\section{A power-law model for $R(L)$}
\label{sec:powerlaw}

Although we can compute the mean flux without specifying the function
$R(L)$, the 
same is not true for a number of other interesting
quantities, such as the variance of the flux and the cross-section
weighted mean luminosity and impact parameter. It is, however,
possible to obtain analytic solutions for these and other quantities if we
make the following additional assumptions:
\begin{enumerate} 
\setcounter{enumi}{6}
\item The luminosity function is a Schechter function.
\item The function $R$ is a power-law of $L$.
\end{enumerate}
If
\begin{equation}
R = R_\ast \left ({L \over L_\ast}\right )^\beta
\label{eq:R}
\end{equation}
(assumption 8), then the normalization factor $R_\ast$ is determined
by the observed rate of incidence:
\begin{equation}
{d{\cal N} \over dz} =
R_\ast^2 \pi f_{\rm cov} \phi_\ast {c \over H} (1+z)^2
\Gamma(1+\alpha+2\beta),
\label{eq:dN/dz_pl}
\end{equation}
which follows from integrating equation (\ref{eq:dN/dz}), and the
flux (eq.~[\ref{eq:F}]) is given by 
\begin{equation}
F  = F_\ast \left ( {L \over L_\ast}\right )^{1-2\beta},
\label{eq:f_pl}
\end{equation}
where
\begin{eqnarray}
F_\ast & \equiv & {L_\ast f_{{\rm esc},N} \over 4\pi R_\ast^2},\\
& = & \left ({d{\cal N} \over dz}\right )^{-1}{c (1+z)^2
\over 4 H} f_{{\rm esc},N}f_{\rm cov} \phi_\ast L_\ast \Gamma(1 + \alpha
+2\beta).
\label{eq:fstar}
\end{eqnarray}
Under these assumptions we can also solve numerically for
the median luminosity and thus for the median flux and impact parameter. 
But before doing so in \S\ref{sec:medians}, let us first consider what a
reasonable value of $\beta$ might be.

It is difficult to predict what the dependence of $R$ on $L$ should
be, which is why it is so convenient that $\bar{F}$ is independent of
this unknown function. In particular, it could well be different for
different elements and for different $N_{\rm min}$. However, a
reasonable guess might be that $R$ 
scales with the virial radius, $R\propto
r_{\rm vir} \propto M_{\rm vir}^{1/3}$, which yields $\beta =
1/[3(1-\gamma)]$ if $M/L \propto M^\gamma$. On the other hand, since $N\sim
n(R)R$ (see \S\ref{sec:assumptions})
which becomes $N\sim n(r_{\rm vir}) r_{\rm
  vir}^2/R$ for a singular isothermal 
profile, it may be that $R\propto r_{\rm vir}^2$ (because $n(r_{\rm
  vir})$ is constant) and that $\beta =
2/[3(1-\gamma)]$ is thus a better guess.

Semi-analytic models of galaxy
formation (e.g., Benson et al.\ 2000) predict that mass-to-light
ratios reach a minimum around $L_\ast$, but that the dependence of $M/L$ on
$M$ is weak. For example, $\gamma \approx 0.1$, in which case we would
expect $\beta \approx 0.37 - 0.74$, is a good fit to the predictions
of Benson et al.\ (2000) (who modeled the $B$-band at $z=0$ for a
$\Lambda$CDM model and a mass range $M \sim 10^{11} - 10^{15}~M_\odot$).

Assuming that $R(L)$ is indeed a power-law, it is possible
to measure $\beta$ by imaging fields containing quasars. Chen et al.\
(2001a, 2001b) have done just this for the $B$-band (the
bluer bands are the most relevant here since we are interested in
ionizing radiation) at $z<1$. They found $\beta = 0.5 \pm 0.1$ for \CIV\ and
$\beta = 0.4 \pm 0.1$ for $N_{\rm HI} > 10^{14}~\cm^{-2}$, in good
agreement with our naive estimates. 

Note that $\beta = 0.5$ is special
as it gives a flux that is independent of luminosity: $F=F_\ast$.
Hence, if $\beta$ is indeed close to
0.5, as both observations and (handwavy) theoretical arguments seem to
suggest, then this has the important implication that \emph{absorbers
belonging to a given class [defined by their rate of incidence $d{\cal
  N}/dz(N>N_{\rm min})$], are nearly all exposed to approximately the
same flux}.  

\subsection{Moments}
\label{sec:moments}

Armed with the fitting functions (\ref{eq:lf}) and (\ref{eq:R}), it
is easy to derive a number of characteristics of the absorbers.
Substituting these functions into equation (\ref{eq:barF}), we obtain the
following expression for the mean flux,
\begin{equation}
\bar{F} = F_\ast {\Gamma(2+\alpha) \over \Gamma(1+\alpha+2\beta)}.
\label{eq:fbar_pl}
\end{equation}
The dependence of $\bar{F}$ on $\beta$ is only apparent, because
$F_\ast \propto R_\ast^{-2} \propto \Gamma(1+\alpha+2\beta)$.
As was shown in \S\ref{sec:meanflux}, the mean flux is independent of
$R(L)$ (and thus $\beta$) when expressed in physical units. 

More generally, it is not difficult to show that the $m$-th moment of
the flux is given by, 
\begin{equation} 
\left < F^m \right > = F_\ast^m
{\Gamma[1+\alpha+m+2\beta(1-m)] \over \Gamma(1+\alpha+2\beta)}.
\end{equation}
Note that $\left <F^m\right > = F_\ast^m$ if $\beta = 0.5$.
Similarly to the moments of the flux, we can derive the moments of
the impact parameter
\begin{equation} 
\left < R^m \right > = R_\ast^m {\Gamma[1+\alpha+\beta(2+m)] \over
  \Gamma(1+\alpha+2\beta)},
\end{equation}
and the luminosity
\begin{equation} 
\left < L^m \right > = L_\ast^m {\Gamma(1+\alpha+2\beta+m) \over
  \Gamma(1+\alpha+2\beta)}.
\end{equation}
Recall that all these moments are weighted by the
cross-section for absorption. Putting in numbers, for
$\beta=0.5$ and $\alpha=-1.6$ (-1.2) we obtain $\bar{L} =
0.9 L_\ast$ ($1.3L_\ast$) and $\bar{R} \approx 
0.5 R_\ast$ ($0.8 R_\ast$). Figure~\ref{fig:meanmed} shows $\bar{F}/F_\ast$
(\emph{left}), $\bar{R}/R_\ast$ (\emph{middle}), and $\bar{L}/L_\ast$
(\emph{right}), all as a function of $\beta$ and for two values
of $\alpha$ as indicated in the figure. 

\subsection{Medians}
\label{sec:medians}

Because $F$ is a monotonic function of $L$ (which is a consequence of
our assumptions 4 (\S\ref{sec:assumptions}) and 8), the median flux
$F_{\rm med} = F(L_{\rm med})$,
and it can be seen from equation (\ref{eq:dN/dz}) that the median
luminosity $L_{\rm med}$
can be obtained by solving the following equation 
\begin{equation}
\int_{L_{\rm min}}^{L_{\rm med}} \Phi R^2 dL 
= {1 \over 2} \int_{L_{\rm min}}^{L_{\rm max}} \Phi R^2 dL.
\label{eq:Lmed}
\end{equation}

\begin{figure*}
\epsscale{1.1} 
\plotone{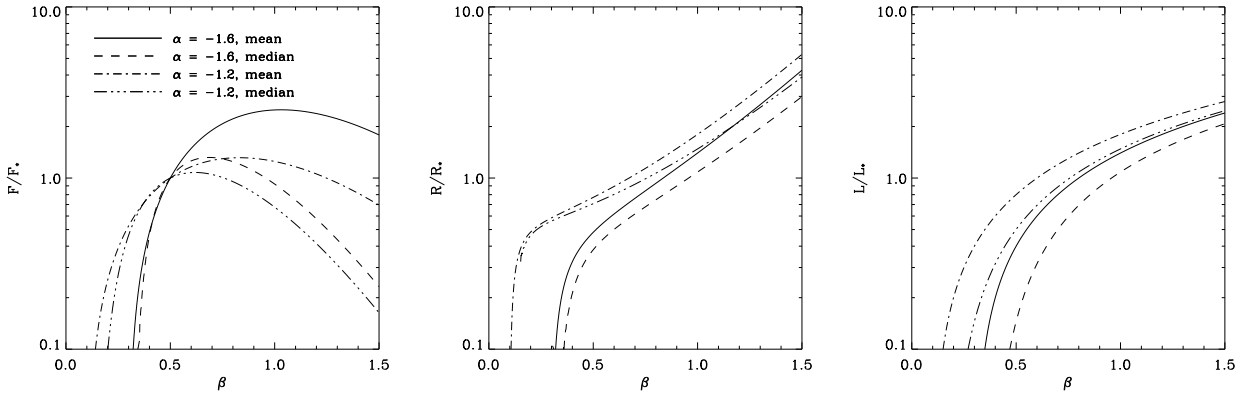}
\figcaption{The flux from the central galaxy seen by the absorbers
  (\emph{left}), the impact 
parameter relative to the central galaxy (\emph{middle}), and the
luminosity of the central galaxy (\emph{right}) are all
plotted as a function of $\beta$, the exponent of the
power-law $R = R_\ast (L/L_\ast)^\beta$. In each panel both the (cross-section
weighted) mean and the median are shown 
for two different slopes of the faint-end of the luminosity function:
$\alpha = -1.6$ and $\alpha = -1.2$. The quantities plotted are all
dimensionless because the flux, impact parameter, and luminosity are plotted
relative to the characteristic quantities $F_\ast$, $R_\ast$, and
$L_\ast$. While the latter is the familiar characteristic luminosity
appearing in the Schechter luminosity function (eq.\
  [\protect\ref{eq:lf}]), the parameters $F_\ast$ and 
$R_\ast$ depend on $\beta$, the rate of incidence of the
absorbers, the parameters of the luminosity function, and the factors
$f_{\rm cov}$ and $f_{{\rm esc},N}$. Their values can be computed 
using equations (\protect\ref{eq:dN/dz_pl}) and
(\protect\ref{eq:fstar}). If the mean flux were plotted in physical
units, rather than relative to $F_\ast$, then it would be independent
of $\beta$. Note that the model predictions become unphysical at small
$\beta$ because $R_\ast\rightarrow 0$ as $\beta \rightarrow
-(1+\alpha)/2$.
\label{fig:meanmed}}
\end{figure*}

It is easy to see from equation (\ref{eq:f_pl}) that for $\beta = 0.5$ we have
$F_{\rm med} = \bar{F} = F_\ast$ because in this case $F$ is
independent of $L$. 
The left-hand panel of Fig.~\ref{fig:meanmed} shows how the median and
the mean flux 
compare for other values of $\beta$ and for two values of the slope of the
faint-end of the luminosity function: $\alpha = -1.6$ and $\alpha =
-1.2$. The results shown in this figure are
insensitive to $L_{\rm min}$ and $L_{\rm max}$ provided that $L_{\rm min}
\ll L_\ast \ll L_{\rm max}$. The mean flux is typically greater
than the median, but they differ by less than a factor of two for
$0.35 < \beta < 0.85$ if $\alpha = -1.6$ ($0.22 < \beta < 1.05$ if
$\alpha = -1.2$). Note that the power-law model for $R(L)$ becomes
unphysical for low $\beta$ because $R_\ast \rightarrow 0$ for $\beta
\rightarrow -(1+\alpha)/2$. 

The median impact parameters are shown in the middle panel of
Fig.~\ref{fig:meanmed}, again for two values of $\alpha$ as indicated
in the figure. As was the case for the flux, the medians are
typically close to the means. 

For the mean flux
to be an order of magnitude greater than the median, we would require $\beta
> 1.6$. Such a large value would be in conflict with the observations
of Chen et al.\ 
(2001a, 2001b) and would probably imply an extremely low value
of $\gamma$ ($\gamma < -1.4$ for $\beta = 2/[3(1-\gamma)]$).
Hence, although the median flux cannot be computed without specifying
$R(L)$, it is within a factor of a few of the mean for plausible
models. This gives us confidence that equation (\ref{eq:barF}) gives a 
reasonable estimate of the typical flux seen by the absorbers. Thus, we
expect that in general $F_{\rm med} \sim \bar{F} \sim F_\ast$.

\section{Application to observations}
\label{sec:observations}

The two types of sources that we will consider are galaxies and
quasars. We will compute the characteristic flux at 
1~Ryd as well as the \HI\ ionization rate for two relatively well
measured luminosity functions: Lyman-break galaxies (LBGs) at $z=3$ (Steidel et
al.\ 1999a) and quasars at $z=2.3$ (Croom et al.\ 2004). 
Both of these luminosity functions were derived using samples selected
in the rest-frame ultraviolet (UV). Hence, they do not include
highly obscured 
sources, which may provide a substantial contribution to the total
rates of star formation and accretion onto black holes in the
universe. However, since such systems would 
likely also be absent from surveys of quasar absorbers, which
typically target quasars that are exceptionally bright in the
rest-frame UV, these luminosity functions are exactly what is needed
for our purposes.

\subsection{Galaxies}
\label{sec:gals}
By combining their ground-based data with data from the Hubble Deep
Field (HDF), Steidel et al.\ (1999a) find that, down to at least $0.1 L_\ast$,
the $z\approx 3$ luminosity function is well fit by a Schechter function 
with parameters $\alpha = -1.60 \pm 0.13$, $\mathcal{R}_{\ast} =
24.48\pm 0.15$ (AB magnitude), and $\phi_\ast = 4.45\times 10^{-3}
h^3\,\Mpc^{-3}$ [for\footnote{Steidel et al.\ (1999a)
    quote $\phi_\ast = 1.6\times 10^{-2}h^3\,\Mpc^{-3}\,{\rm
      mag}^{-1}$ for $(\Omega_m,\Omega_\Lambda) = (1.0,0.0)$. We
changed the units from per magnitude to per $\ln L$ by dividing by
$0.4\ln 10$ and we 
    converted to $(\Omega_m,\Omega_\Lambda) =
(0.3,0.7)$ using the effective volumes listed in their Table~3.}
  $(\Omega_m,\Omega_\Lambda) = (0.3,0.7)$ and $\left <z \right> =
3.04$]. We note that because they relied
on HDF data for $L < L_\ast$, which covers only a relatively small
field, the faint-end slope may be subject to
systematic errors due to cosmic variance.

$\mathcal{R}_{\ast}(AB) = 24.48\pm 0.15$ at $z =
3.04$ yields $L_{\nu,\ast} = (5.72 \pm 0.79) \times
10^{28}~h^{-2}~\erg\,\s^{-1}\,\Hz^{-1}~f_{\nu,{\rm esc}}^{-1}$ %5.7174756 
at rest-frame $\lambda = \lambda_{\rm eff}/(1+z) = 1715.35$~\AA. 
Starburst99 (version 4.0, Leitherer et al.\ 1999) predicts
$L_\nu(\lambda 1715)/L_\nu(\lambda 912) = 3.6$ for a
continuous starburst (the result converges within 10~Myr)
with a Salpeter initial mass function with upper and lower mass cutoffs
of $100~M_\odot$ and $1~M_\odot$ respectively, and a metallicity $Z =
0.2Z_\odot$ [$L_\nu(\lambda 1715)/L_\nu(\lambda 912) = 4.1$ for $Z =
Z_\odot$].  
Using this value we obtain $L_{\nu,\ast}(\lambda 912) = (1.59 \pm 0.22)\times
10^{28}~h^{-2}~\erg\,\s^{-1}\,\Hz^{-1}~f_{{\rm esc},\lambda1715}^{-1}$.
Equation (\ref{eq:barF}) then gives
\begin{equation}
{\bar F}_\nu (\lambda 912)_{{\rm LBG},z=3} = (4.43 \pm 0.90)
\times 10^{-20}~\erg\,\cm^{-2}\,\s^{-1}\,\Hz^{-1}
~ f_{\rm cov} {f_{{\rm
      esc},N,\lambda912} \over f_{{\rm esc},\lambda1715}} \left
({d{\cal N} \over 
  dz}\right )^{-1}.
\label{eq:barF_LBG}
\end{equation}%4.4343629e-20
Using the same Starburst99 model for the spectral shape, we find that
this corresponds to a \HI\ ionization rate of  
\begin{equation}
\bar{\Gamma}_{{\rm HI,LBG},z=3} = (9.94 \pm 2.03)\times
10^{-12}~\s^{-1}~f_{\rm cov} {\left < f_{{\rm esc},N}\right > \over f_{{\rm
      esc},\lambda1715}} \left ({d{\cal N} \over dz}\right )^{-1},
\label{eq:Gamma_LBG}
\end{equation}%9.9439024e-12 
where $\left <f_{{\rm esc},N}\right >$ is the average fraction
of hydrogen ionizing photons that escapes to the absorber, weighted by
the cross-section for ionization. 
Neglecting evolution in the luminosity density, the values for
$z=2.3$ would be 0.90 times those for $z=3$.

It is interesting to estimate the typical impact parameter of
absorbers around LBGs. Assuming the power-law radius-luminosity
relation (eq.~[\ref{eq:R}]), equation (\ref{eq:dN/dz_pl}) yields,
\begin{equation}
R_{\ast,{\rm LBG},z=3} = 1.16 \times 10^2~\kpc~\left ( {d{\cal N} \over dz}
  \right )^{1/2}\left [f_{\rm 
  cov}\Gamma(2\beta-0.4)\right ]^{-1/2}, %116490.23
\label{eq:rstar}
\end{equation}
where we assumed $h=0.7$.
Using $\beta = 0.5$ gives $\bar{R} \approx 48~\kpc~ (d{\cal
    N}/dz)^{1/2} f_{\rm cov}^{-1/2}$.
To put this in context, damped \lya\ systems (DLAs, $N_{\rm
HI,min} = 2\times 10^{20}~\cm^{-2}$) have an observed rate of
incidence $d{\cal N}/dz = 0.20$ at 
$z\approx 3$ (Storrie-Lombardi \& Wolfe 2000). Hence, DLAs are likely
to have impact parameters of $\sim 10$~kpc (see also Fynbo et al.\
1999; Haehnelt, Steinmetz, \& Rauch 2000).
Comparing this with the half-light ratios of bright ($L\ga L_\ast$)
LBGs, which are typically about 2~kpc
(0.2--0.3~arcsec; Giavalisco Steidel, \& Macchetto 1996) we see that
the approximation of point-like sources is likely to be reasonable.
However, we do expect our model to overestimate the local flux for
absorbers that are much rarer than DLAs. 

We can also use equation (\ref{eq:rstar}) to verify the validity of
our assumption that the flux from nearby sources is negligible
compared to that from the central source.
Since galaxies with $L\sim L_\ast$ dominate the cosmic luminosity
density, we would expect this assumption to
be a good approximation if $(1+z)^3R_\ast^3\phi_\ast \ll 1$ if
the galaxies were randomly distributed. From equation
(\ref{eq:rstar}) we can see that this would imply $d{\cal 
  N}/dz \ll 4 \times 10^2 f_{\rm cov}\Gamma(2\beta-0.4)$. However,
bright LBGs are observed to cluster: their inferred comoving
correlation length is $r_0 = 5h^{-1}\,\Mpc$ for a two-point correlation
function of the form $\xi = (r/r_0)^{-1.8}$ (Steidel et al.\
1999b). Hence, the expected 
number of galaxies within a distance $R_\ast$ of a $L_\ast$-galaxy is
about $2.5[R_\ast(1+z)/r_0]^{-1.8}$ higher than for a random
distribution and we require $d{\cal
  N}/dz \ll 1 \times 10^2 f_{\rm cov}\Gamma(2\beta-0.4)$ for the
flux from nearby galaxies to be negligible. We
will see in \S\ref{sec:comparison} that for $z \le 4$ this condition is
just satisfied if 
the flux (at around 1~Ryd) from the central galaxy exceeds the
background. However, given the approximate nature of this estimate, it
is possible that we have underestimated the mean local flux by a
factor of a few for classes of absorbers for which the flux from the
central galaxy is not large compared with the background. 

There is one class of absorbers for which the intensity of the local
UV radiation has been measured: DLAs with detectable absorption by
\CII$^\ast$. Wolfe et al.\ (2004) have measured the UV flux at
1500~\AA\ (rest-frame) for the 23 absorbers in their sample of 45
$z\sim 2$--4 DLAs for which they were able to obtain a positive
detection of \CII$^\ast$ absorption.  By equating inferred cooling
rates to grain photo-electric heating rates (given the measured
dust-to-gas ratios), they find that their \CII$^\ast$ absorbers are
typically exposed to a flux that is within a factor of a few of $F_\nu
\sim 4\pi 10^{-18.5}\approx 4\times
10^{-18}~\erg\,\cm^{-2}\,\s^{-1}\,\Hz^{-1}$. %3.97384e-18 
Our Starburst99 spectral template predicts $F_\nu(1500)/F_\nu(1715) =
1.12$, which yields 
$F_\nu (\lambda1500)= 2\times 10^{-18}~
\erg\,\cm^{-2}\,\s^{-1}\,\Hz^{-1} ~f_{\rm cov} f_{{\rm
    esc},N,\lambda1500}/f_{{\rm esc},\lambda1715}$ for $d{\cal N}/dz =
0.1$ ($\approx 
0.20 \cdot 23/45$, where 0.20 is the rate of incidence for $z\approx
3$ DLAs measured by Storrie-Lombardi \& Wolfe 2000). %1.7856423e-18
Thus, \emph{for the one class of absorbers for which the local UV flux has
been measured, the prediction of our model is in excellent agreement
with the observations}. Measurements of the local flux based on ${\rm
  H}_2$ lines confirm that the UV radiation field in DLAs far exceeds the
background (e.g., Hirashita \& Ferrara 2005, Srianand et al.\ 2005).

The good agreement between our prediction for the flux to which DLAs
are exposed, which relied on the assumption that DLAs arise in the
halos of LBGs, and the measurement of Wolfe et al.\ (2004) suggests
that DLAs and LBGs may be drawn from the same population of
galaxies, as has already been suggested on other grounds by
Schaye (2001b) and M{\o}ller et al.\ (2002). If this is indeed the case, then
we can read of the predictions of our model for the impact parameters
and luminosities from figure~\ref{fig:meanmed}. For $\beta\approx 0.5$
and $z=3$ we predict a median impact parameter ${\rm med}(R) \approx
0.4 R_\ast \approx 14~\kpc \approx 1.8$~arcsec) (see
  eq.~[\ref{eq:rstar}]) and a median 
luminosity ${\rm med}(L) \approx 0.1 L_\ast$.

\subsection{Quasars}
\label{sec:quasars}

Using the double power-law fit to the luminosity function of quasars
in the 2dF QSO redshift survey (Croom et al.\ 2004), we find  $\int
L\Phi(L)dL = 2.67\times 10^{-6}L_\ast ~\Mpc^{-3}$ and $M_{b_{\rm J},\ast} =
-25.77$ for $(\Omega_m,\Omega_\Lambda,h) = (0.3,0.7,0.7)$ and
$z=2.3$. Assuming $b_{\rm J} = B$, a power-law spectrum $F_\nu =
F_{\nu}(\nu_{\rm LL}) (\nu/\nu_{\rm LL})^{-\alpha_\nu}$, 
where $\nu_{\rm LL}$ is the frequency at the hydrogen Lyman limit, and a
spectral index $\alpha_\nu = 1.8$ (Telfer et al.\ 2002)
this gives $L_{\nu,\ast}(\lambda 912) = 5.23\times
10^{30}~\erg\,\s^{-1}\,\Hz^{-1}~f_{{\rm esc},\lambda1333}^{-1}$ and
thus a mean flux
\begin{equation}
{\bar F}_\nu (\lambda 912)_{{\rm QSO},z=2.3} = 5.05
\times 10^{-21}~\erg\,\cm^{-2}\,\s^{-1}\,\Hz^{-1}
~ f_{\rm cov} {f_{{\rm esc},N,\lambda912}\over f_{\rm
    esc,\lambda1333}} \left ({d{\cal N} \over
  dz}\right )^{-1}.
\end{equation}%5.0471172e-21
For a power-law spectrum the \HI\ ionization rate is $\Gamma_{\rm HI}
= 9.51\times 
10^8~F_\nu(\nu_{\rm LL})/(3+\alpha_\nu)~\s^{-1}$. Hence, if there
exists a population of absorbers which all reside in halos around
QSOs, then such absorbers will on average be exposed to a local
radiation field with an ionization rate
\begin{equation}
\bar{\Gamma}_{{\rm HI,QSO},z=2.3} = 1.00\times
10^{-12}~\s^{-1}~f_{\rm cov} {\left < f_{{\rm esc},N} \right > \over f_{\rm
    esc,\lambda1333}} \left 
({d{\cal N} \over dz}\right )^{-1}.  
\end{equation}%1.0005312e-12

Since both $\bar{F}$ and the intensity of the UV background (UVB)
radiation are proportional to the luminosity density of the sources,
the relative contributions of the central 
galaxies and quasars to $\bar{F}$ are identical to their relative
contributions to the UVB.

However, although every quasar probably resides in a galaxy, only a
few percent of galaxies host an active quasar (e.g., Steidel et al.\
2002). Therefore, the mean flux from local quasars may not be very
relevant to quasar absorption studies. 
Studies employing non-parametric statistics (such as the
median and other percentiles) and studies using small samples of absorbers, are
unlikely to be 
significantly affected by local quasar radiation, unless the absorbing
species requires the presence of a hard radiation source for its
existence. It should, however, be noted that regardless of how small
the fraction 
of time that a quasar is active, local radiation from quasars would be
important if they are dormant for periods comparable to or smaller
than the timescale for the establishment of ionization
equilibrium. Although this timescale is typically very short for
hydrogen ($\la 10^5~\yr$), it can be much longer for species with higher
ionization potentials.

In the remainder of this paper we will ignore radiation from local 
quasars. 

\subsection{Comparison with the background}
\label{sec:comparison}

How does the flux from local sources compare to that from the
extragalactic background? In section \ref{sec:comp} we found that,
relative to the background, the local flux is given by
\begin{equation}
{\bar{F} \over F_{\rm bg}} = \left ({d{\cal N} \over dz}\right )_{\rm
  crit} \left({d{\cal N} \over dz}\right )^{-1},
\end{equation}
where the critical rate of incidence can be computed using either
equation (\ref{eq:dndzcrit1}) or (\ref{eq:dndzcrit2}). The former
method requires a measurement of the luminosity density, the
background flux, and the escape
fraction to the absorber ($f_{{\rm esc},N}$), while the latter
requires measurements of 
the mean free path and the ratio $f_{{\rm esc},N}/f_{\rm esc}$, and
also requires one to assume that the sources associated with the absorbers
dominate the background. As the
absorber column density tends to the 
column in front of the sources, $f_{{\rm esc},N}$ asymptotes to
unity, while for much lower absorber column 
densities it asymptotes to the global escape fraction $f_{\rm
  esc}$. Hence, provided the mean free path is known and that the
sources associated with the absorbers dominate the background, the second
method is probably more reliable for all but the highest column
densities. 
If the luminosity density, the background, and the mean free path are
all known, then we can measure the global escape fraction $f_{\rm
  esc}$ by comparing the two methods.

We will now apply both methods to hydrogen ionizing photons. 
For other ions the calculation is entirely analogous to that
for \HI,
except that a model of the spectral shape of the UVB 
is required because the rates of photo-ionization 
of heavier elements by the background have not been measured.
For ions with ionization potentials $E_{\rm ion} < 4~\Ryd$ we expect
the critical rates of incidence to be similar to those of \HI, but for
ions with $E_{\rm ion} > 4~\Ryd$, such as \OVI, the critical rates
should be much smaller because stars (with the exception of Pop.\ III)
emit very little radiation at these energies. Note, however, that even
though local 
stellar radiation is unlikely to be important above 4~Ryd, ionization
models of \OVI\ absorbers (as well as other ions with ionization
potentials above 4~Ryd) generally also use constraints from other
species whose ionization balance is affected by the addition of
radiation below 4~Ryd. 

First, we will compute the critical rate of incidence using method 1
(i.e., equation \ref{eq:dndzcrit1}). For the intensity of the UV
background we use $\Gamma_{\rm HI}/(10^{-12}~\s^{-1}) = 
(0.06,0.67,1.3,0.9,1.0)$ at $z=(0,1,2,3,4)$. These
measurements were taken from\footnote{Where appropriate the
measurements have been scaled to the currently favored cosmology
($\Omega_bh^2=0.0224$, $h=0.71$; Spergel et al.\ 2003) using the
relation $\Gamma \propto \Omega_b^2h^3$ (e.g., Rauch et al.\ 1997).}
Dav\'e \& Tripp (2001) ($z=0$), Scott et 
al.\ (2002) ($z=1$), and Bolton et al.\ (2005) ($z=2$--4). For
redshift 3 we have already computed the mean ionization rate due to
local sources (equation \ref{eq:Gamma_LBG}). We can scale this result
to other redshifts if we assume that the luminosity density of ionizing 
radiation is proportional to the cosmological star formation density, for
which we take $\dot{\rho}_\ast(z=0) = 0.1 \dot{\rho}_\ast(z=1)$ and
$\dot{\rho}_\ast(z>1) = \dot{\rho}_\ast(z=1)$ (e.g., Heavens
et al.\ 2004 and references therein). We then find that
\begin{equation} 
\left ({d{\cal N} \over dz}\right )_{\rm crit} =
(5,\,9,\,6,\,11,\,11)\,{\left < f_{{\rm esc},N}\right >f_{\rm cov}
  \over f_{{\rm 
      esc},\lambda1715,z=3}} ~~~~{\rm at}~~ 
z=(0,\,1,\,2,\,3,\,4).
\label{eq:resulting_dndz1}
\end{equation}

Second, we compute the critical rate of incidence using method 2,
which reduces to equation (\ref{eq:dndzcrit3}) for \HI, if we
neglect the variation of the mean free path with frequency and if we
assume that the \HI\ column density distribution is of the form 
$d{\cal N}/dN \propto N_{\rm HI}^{-1.5}$ around the Lyman
limit. Making this assumption and using $\left (d{\cal N}/dz\right
)_{\rm LL} \approx (0.7,1.0,1.3,2.0,4.0)$ for $z = (0,1,2,3,4)$
(P\'eroux et al.\ 2005), we obtain
\begin{equation}
\left ({d{\cal N} \over dz}\right )_{\rm crit} =
(0.3,\,0.4,\,0.6,\,0.9,\,1.8)\,{\left <f_{{\rm esc},N} \right >f_{\rm
    cov} \over f_{\rm esc}} ~~~~{\rm at}~~  
z=(0,\,1,\,2,\,3,\,4). 
\label{eq:resulting_dndz2}
\end{equation}
We stress that the ratio of $f$-factors could be greater than unity
for column densities similar to those shielding the sources.

Comparing equations (\ref{eq:resulting_dndz1}) and
(\ref{eq:resulting_dndz2}), we find 
\begin{equation}
{f_{\rm esc} \over f_{{\rm esc},\lambda1715,z=3}} = 
(0.06,\,0.05,\,0.09,\,0.08,\,0.16) ~~~~{\rm at}~~  
z=(0,\,1,\,2,\,3,\,4).
\label{eq:fesc}
\end{equation}
These measurements of the global escape fractions result from the
demand that the luminosity density due to LBGs gives rise to the
measured UV background. Because the measurements of the luminosity
density, the mean free path, and the intensity of the UV background
are all uncertain by a factor of a few or more, these measurements
should be taken as order of magnitude estimates only. Thus, we
conclude that a global escape fraction $f_{\rm esc}\sim 10^{-1}f_{{\rm
esc},\lambda1715,z=3}$ is consistent with all the measurements. If
  quasars dominate the UVB then the escape fraction should be smaller.

Steidel, Pettini, \&
Adelberger (2001) measured $F_\nu(\lambda 1500) / F_\nu(\lambda 900) = 
4.6\pm 1.0$ from their composite LBG spectrum (after correction for
intervening absorption) whereas our spectral template predicts an
intrinsic value of $L_\nu(\lambda 1500) / L_\nu(\lambda 900) = 4.0$,
implying that $f_{\rm esc} \sim f_{{\rm
    esc},\lambda1500,z=3}$. Although this may be consistent with
equation (\ref{eq:fesc}) within the errors (which are very poorly
constrained) there is some tension between the two. However, since the galaxies
studied by Steidel et al.\ (2001) were unusually blue, it is important
to keep in mind that their measurements may not be representative for
the population as a whole.

If $\left <f_{{\rm esc},N}\right >f_{\rm cov} \sim f_{\rm esc}$, then
local ionizing radiation becomes important for \HI\ for LL
systems. Absorbers with $N_{\rm HI} > 10^{19}\,\cm^{-2}$ have rates
of incidence much lower than LL systems (e.g., P\'eroux et al.\ 2005)
and are thus dominated by local radiation.  
Since estimates of the local UV flux have
generally not been available, the implication is that published
results from ionization models of such systems may be significantly in
error. Note that for $N_{\rm HI} > 10^{20}\,\cm^{-2}$ the absorber
columns are probably comparable to those in front of the sources in
which case $f_{{\rm esc},N}$ may be much greater than $f_{\rm esc}$,
further boosting the local flux.

Comparison of the rates of incidence of metal line absorbers
with the critical rates indicates that UV
radiation from local galaxies may not be negligible
for strong metal systems, such as those containing
detectable absorption by ions in low ionization stages. For example,
using the data from Boksenberg et al.\ (2003), 
who decomposed a large number of high-quality quasar absorption
spectra into Voigt profiles, we find\footnote{The columns quoted are
integrated over the systems (i.e., the columns of the
individual Voigt profile components were summed) and the rates of
incidence are for systems at redshift $z=2.5$--3.5. We only used data
with observed wavelength redwards of the quasar's \lya\ emission line
and with a redshift more than $4000~\kms$ bluewards of the redshift of
the quasar.} $d{\cal N}/dz = 4.5$ for $\log N_{\rm min}($\CII$) =
12$ and $d{\cal N}/dz = 3.4$ for $\log N_{\rm min}($\SiIV$)
= 13$, both at $z\approx 3$. These observed 
rates of incidence are likely to be overestimates of the true
rates since many of the quasars in the sample of Boksenberg et al.\
(2003) were originally selected to contain DLAs. Indeed, more than
half of their \CII\ systems with $\log N>12$ and nearly half of their
\SiIV\ systems with $\log N > 13$ are associated with DLAs.
Other frequently studied metal line systems with low
rates of incidence include \MgII\ systems ($d{\cal
  N}/dz\approx 1$ for systems with an equivalent width ${\rm
  W}_\lambda > 0.3$~\AA\ at 
$z\sim 1$; Steidel \& Sargent 1992). 

It is difficult to predict exactly how the physical properties
inferred from ionization models would change if local radiation had
been taken into account, because the effects will depend on the ions
involved as well as on the rate of incidence of the absorbers. However,
if the column density of \HI\ is used as a constraint, then 
the addition of local radiation will generally tend to increase the
inferred density and to decrease the inferred size of
the absorber. The 
effects on the abundances of heavy elements are more difficult to
predict and will likely be different for different elements.

\section{Conclusions}
\label{sec:conclusions}

We have constructed an analytic model to estimate the
characteristic flux from local sources of radiation to which quasar
absorption systems are exposed. Since many studies have shown that
fluctuations in the UV background are unimportant for the low-column
density \HI\ \lya\ forest (at least at $z < 5$), we focused on the rarer
systems with higher \HI\ columns and/or detectable absorption by
heavy elements. 

The most important assumptions we made are that the absorption arises
in a roughly 
spherical gas cloud centered on a source of ionizing radiation, that
radiation from other nearby sources is negligible, and
that the
probability that a sightline intersects a total column density $N> N_{\rm
  min}$ is zero beyond some radius $R$ (which may depend on the
luminosity $L$ and $N_{\rm min}$). The first assumption will break
down for the low column 
density \lya\ forest, which is thought to arise in a web of 
sheet-like and filamentary structures, but is likely to be reasonable
for the rarer systems with higher \HI\ columns and/or detectable
absorption by heavy elements. The second assumption is
conservative. We argued it is likely to be a reasonable approximation for
those absorbers for which we predict the flux from
the central galaxy to exceed
the background (around 1~Ryd). The third assumption implies that
higher columns arise in sightlines with smaller impact parameters,
which we argued must typically be the case. 

We showed that the mean flux from the central galaxy is 
proportional to the luminosity density of the sources and inversely
proportional to the rate of incidence $d{\cal N}/dz(N>N_{\rm min})$ of
the absorbers, but that it does 
\emph{not} depend on the function $R(L,N_{\rm min})$. Assuming a
power-law dependence $R\propto L^\beta$, we derived analytical
expressions for the cross-section weighted moments of the flux,
of the impact parameter, and of the luminosity. In addition, we computed the
corresponding medians numerically. We found that for
reasonable values of $\beta$, the distribution of fluxes is narrow and
the mean (which is independent of $\beta$) is of the same order of
magnitude as the median. For
the special case of $\beta=0.5$ all absorbers belonging to a class
with a given rate of incidence are exposed to the same
flux. Interestingly, both observations and naive theoretical arguments
suggest that $\beta$ is close to this value. This implies that
there exists a characteristic flux to which absorbers with $N>N_{\rm
  min}$ are exposed,
which can be estimated using equation (\ref{eq:barF}).  

We applied our model to two relatively well-studied sources of ionizing
radiation: Lyman-break galaxies (LBGs) at $z\approx 3$ and quasars at
$z\approx 2$. We argued that galaxies are more relevant to quasar
absorption systems because quasars are thought to be dormant for most
of their lifetimes.

Our predictions are in 
excellent agreement with the observations for the one class of
absorbers for which the flux has been measured: 
DLAs at $z\approx 3$ with detectable absorption by \CII$^\ast$,  Since
our calculation 
assumed that the absorbers are centered on LBGs, this agreement
suggests that DLAs and LBGs may be drawn from the same underlying
population of galaxies. We predicted (for $\beta\approx 0.5$) that DLAs
at $z\approx 3$ typically have impact 
parameters of order 10~kpc and that the median luminosity of their
LBG counterparts is about $0.1L_\ast$. Predictions for classes of
absorbers with other rates of incidence can easily be obtained from
the expressions derived in \S\S\ref{sec:method} and \ref{sec:powerlaw}.

We used two different methods to compare the mean, local flux to the
background. The first takes the luminosity density and the escape
fraction to the absorber as inputs, and requires
knowledge of the background intensity. The second takes the mean free
path for photons and the ratio of the escape fraction to the absorber
and the global escape fraction as inputs, and relies on the assumption
that the sources associated with the absorbers dominate the
background. We found that consistency between the two methods, which
comes down to the requirement that the observed sources make up the
observed background, requires a global escape fraction of order 10
percent. 

For absorbers as rare or rarer than Lyman limit systems (which have
$d{\cal N}/dz \approx 2$ at $z=3$), local H ionizing radiation becomes
important. (Sub-)DLA systems are likely dominated by local
radiation. Studies that have modeled the ionization balance of such
systems assuming that they are exposed to the background only, may
therefore have produced spurious results.

\acknowledgments
It is a pleasure to thank Jordi Miralda-Escud\'e for enlightening
discussions which greatly improved this work and Anthony Aguirre
for a careful reading of the 
manuscript. I also thank the Institute for Computational Cosmology in
Durham and the Institute of Astronomy in Cambridge, where part of this
work was completed. Finally, I gratefully acknowledge support from the
W.~M.~Keck Foundation, from NSF grant PHY-0070928, and from Marie
Curie Excellence Grant MEXT-CT-2004-014112.

\end{document}